\documentclass[floats,floatfix,amssymb,prd,twocolumn,superscriptaddress,nofootinbib,groupedaddress]{revtex4-1}
	
\usepackage{subcaption}
\usepackage{ragged2e}
\DeclareCaptionJustification{justified}{\justifying}
\usepackage{amssymb,amsmath,verbatim,mathtools,needspace,enumitem,etoolbox,graphicx,physics,microtype,afterpage,bm}
\usepackage{esint}
\usepackage[dvipsnames]{xcolor}
\definecolor{linkcolor}{rgb}{0.0,0.3,0.5}
\usepackage[unicode, colorlinks=true, linkcolor=linkcolor, citecolor=linkcolor, filecolor=linkcolor,urlcolor=linkcolor, pdfusetitle]{hyperref}
\usepackage[all]{hypcap}
\usepackage[T1]{fontenc}
\usepackage[utf8]{inputenc}
\usepackage{tabularx}
\usepackage{float}
\newcommand\underrel[3][]{\mathrel{\mathop{#3}\limits_{%
      \ifx c#1\relax\mathclap{#2}\else#2\fi}}}
\usepackage{soul}
\usepackage{lmodern}
\allowdisplaybreaks
\usepackage{tikz}
\usepackage{cleveref}
\usepackage{color}
\usepackage{framed}
\usepackage{hyperref}
\hypersetup{colorlinks, citecolor=bluscuro, linkcolor=black, urlcolor=bluscuro}
\definecolor{rossos}{cmyk}{0,1,1,0.55}
\definecolor{bluscuro}{rgb}{0.15, 0.2, .85}

\newcommand{\be}{\begin{equation}}
\newcommand{\ee}{\end{equation}}

\def\lsim{\mathrel{\rlap{\lower4pt\hbox{\hskip0.5pt$\sim$}}
    \raise1pt\hbox{$<$}}}         
\def\gsim{\mathrel{\rlap{\lower4pt\hbox{\hskip0.5pt$\sim$}}
    \raise1pt\hbox{$>$}}}         

\makeatletter
\newcommand{\subsetsim}{\mathrel{\mathpalette\subset@sim\relax}}
\newcommand{\subset@sim}[2]{%
  \vtop{\offinterlineskip\m@th
    \ialign{\hfil##\cr
     ~$#1\subset$\cr\noalign{\kern0.5pt}\scalebox{0.9}{$#1\sim$}\cr
    }%
  }%
}
\makeatother
\makeatletter
\def\l@subsubsection#1#2{}
\makeatother

\begin{document}

\title{Tidal Love numbers of analogue black holes}

\author{Valerio De Luca}
\email{vdeluca@sas.upenn.edu}

\author{Brandon Khek}
\email{bnkhek@sas.upenn.edu}

\author{Justin Khoury}
\email{jkhoury@upenn.edu}

\author{Mark Trodden}
\email{trodden@upenn.edu}

\affiliation{\vspace{0.1cm} Center for Particle Cosmology, Department of Physics and Astronomy,
University of Pennsylvania 209 South 33rd Street, Philadelphia, Pennsylvania 19104, USA}


\begin{abstract}
\noindent
Tidal Love numbers quantify the conservative static response of compact objects to external tidal fields, and are found to vanish exactly for asymptotically flat black holes in four-dimensional general relativity. Many aspects of the physics of black holes have an analogue in the theory of supersonic acoustic flows, including the existence of an event horizon and associated phenomena, such as quasinormal modes and superradiance. In this paper, we investigate the tidal Love numbers of acoustic black holes in different number of dimensions. We find that they exhibit a number of similar properties as higher-dimensional general relativistic black holes, such as logarithmic running with radial distance, and vanishing tidal response for special multipole moments. We show that the latter is a consequence of ladder symmetries, analogous to those identified for black holes. 

\end{abstract}

\maketitle

\section{Introduction}
\label{sec:intro}
\noindent
Tidal effects play a critical role in the inspiral of binary systems. They are usually parametrized in terms of a set of coefficients called tidal Love numbers (TLNs), describing the linear conservative response of self-gravitating bodies to external tidal fields~\cite{1909MNRAS..69..476L}. Initially formulated within the framework of Newtonian gravity, TLNs have been generalized to a fully relativistic context~\cite{Hinderer:2007mb, Binnington:2009bb, Damour:2009vw}, and are encoded in the gravitational waveform of two-body systems. Upcoming data from gravitational wave experiments, including the next generation LISA, Einstein Telescope and Cosmic Explorer~\cite{Punturo:2010zz, Sathyaprakash:2019yqt, Maggiore:2019uih, Reitze:2019iox, Kalogera:2021bya,Branchesi:2023mws,LISA:2017pwj,Colpi:2024xhw}, and the accurate modeling of tidal interactions, promise to offer valuable insights into the interior structure of compact objects. These include, for instance, information on the equation of state of neutron stars~\cite{GuerraChaves:2019foa, Chatziioannou:2020pqz}, or the presence of new physics at the event horizon of black holes~\cite{Maselli:2018fay, Datta:2021hvm}.

The static TLNs for different families of asymptotically flat black holes in four-dimensional general relativity (GR) are known to be exactly zero~\cite{Deruelle:1984hq,Binnington:2009bb, Damour:2009vw, Damour:2009va, Pani:2015hfa, Pani:2015nua, Gurlebeck:2015xpa, Porto:2016zng, LeTiec:2020spy, Chia:2020yla, LeTiec:2020bos, Hui:2020xxx, Charalambous:2021mea, Charalambous:2021kcz, Creci:2021rkz, Bonelli:2021uvf, Ivanov:2022hlo, Charalambous:2022rre, Katagiri:2022vyz, Ivanov:2022qqt,BenAchour:2022uqo, Berens:2022ebl, Bhatt:2023zsy, Sharma:2024hlz, Rai:2024lho}. This property is violated, however, in scenarios involving BH mimickers and exotic compact objects~\cite{Pani:2015tga, Cardoso:2017cfl,Herdeiro:2020kba,Chen:2023vet,Berti:2024moe}, in the presence of a cosmological constant~\cite{Nair:2024mya, Franzin:2024cah}, or
with matter in the environment surrounding the objects~\cite{Baumann:2018vus, DeLuca:2021ite, DeLuca:2022xlz, Brito:2023pyl, Capuano:2024qhv,Cardoso:2019upw, Cardoso:2021wlq, Katagiri:2023yzm, DeLuca:2024uju, Cannizzaro:2024fpz}. It is similarly violated in theories of modified gravity~\cite{Cardoso:2017cfl, Cardoso:2018ptl, DeLuca:2022tkm, Barura:2024uog} and in higher dimensions~\cite{Kol:2011vg, Cardoso:2019vof, Hui:2020xxx, Rodriguez:2023xjd, Charalambous:2023jgq, Charalambous:2024tdj, Charalambous:2024gpf, Ma:2024few}. Recently, TLNs have been generalized to include the effect of non-linearities and for time-dependent tidal perturbations~\cite{DeLuca:2023mio, Riva:2023rcm, Ivanov:2024sds, Iteanu:2024dvx, Kehagias:2024rtz, Combaluzier-Szteinsznaider:2024sgb, 
Nair:2022xfm, Saketh:2023bul, Perry:2023wmm, Chakraborty:2023zed, DeLuca:2024ufn, Katagiri:2024wbg, Katagiri:2024fpn}.

Despite the huge progress made in understanding black holes, both theoretically and with gravitational wave experiments, a complete understanding of their properties is still lacking. This has motivated the development of analogue systems, in particular acoustic black holes (ABHs), and models of analogue gravity. See Refs.~\cite{Visser:1997ux,Barcelo:2005fc} for reviews. As shown by Unruh~\cite{Unruh:1980cg}, a map can be drawn between certain aspects of the physics of black holes and the theory of supersonic acoustic flows. If a fluid flows through a tube with varying cross-sectional area, its velocity will increase whenever it travels through a narrower section of the tube, and may thus exceed the local sound speed. This phenomenon leads to the appearance of an acoustic horizon for sound waves, in analogy to the event horizon of black holes in GR. 
Acoustic black holes share other properties with their gravitational counterparts, including geodesics, wave phenomena like quasinormal modes, superradiance and tail effects, and even Hawking radiation~\cite{Basak:2002aw, Berti:2004ju, Cardoso:2004fi, Cardoso:2005ij, Lepe:2004kv, Kim:2004sf, Saavedra:2005ug, Abdalla:2007dz,Vieira:2014rva, Vieira:2021xqw, Vieira:2021ozg, Singh:2024qfw}.

The main purpose of this work is to point out a further correspondence between black holes and ABHs based on their tidal response. By studying the excitation of acoustic disturbances in the flowing fluid induced by an external tidal field, we compute the TLNs of ABHs in 2+1 and in 3+1 dimensions, where in the latter case we consider both the canonical and de Laval nozzle metrics. Because the acoustic metric differs from the Schwarzschild/Kerr geometry in four dimensions, we find, perhaps unsurprisingly, that acoustic TLNs are in general non-vanishing. Nevertheless, the tidal response of ABHs reproduces a number of properties of higher-dimensional black holes, such as logarithmic running with radial distance in the de Laval nozzle case, as well as the vanishing of TLNs only for certain angular multipole moments. As we will show, the latter property can be explained by the existence of a ladder symmetry in the perturbation equations, thereby exhibiting a strong similarity with GR black holes~\cite{Hui:2021vcv,Berens:2022ebl,Rai:2024lho, Combaluzier-Szteinsznaider:2024sgb}. 

The outline of the paper is as follows. In Sec.~\ref{sec: TLN-GR} we describe the general framework to compute TLNs, studying the example case of Schwarzschild black holes in arbitrary dimensions.  In Sec.~\ref{sec: ABH} we review the correspondence between black holes and models of analogue gravity. In Secs.~\ref{sec: 2+1} and \ref{sec: 3+1} we compute the static Love numbers of ABHs in 2+1 and 3+1 dimensions, respectively. In Sec.~\ref{sec: Ladder}, we identify a ladder symmetry of the perturbation equations, which underlies the vanishing of TLNs at special multipole values. The conclusions are left to Sec.~\ref{conclusions}. We use geometrical units~$G = c = 1$, and mostly positive metric signature.

\section{Tidal Love numbers}
\label{sec: TLN-GR}
\noindent
We first review the formalism for extracting TLNs in the context of Newtonian gravity and GR.

\subsection{Newtonian gravity}
\noindent
Consider a spherically symmetric, rotating body of mass~$M$, located at the origin of a Cartesian coordinate system.
The body will experience tidal deformations when subjected to an external tidal gravitational field,~$U_\text{\tiny ext}$.
Given the spherical symmetry of the unperturbed configuration, it is convenient to decompose~$U_\text{\tiny ext}$ in terms of multipole moments,\footnote{As such,~$U_\text{\tiny ext}$ solves Laplace's equation,~$\vec{\nabla}^2 U_\text{\tiny ext}  =0$ with smooth boundary condition at $r = 0$.} 
\be
\label{eq:UMultipole}
U_\text{\tiny ext} =-\sum_{\ell = 2}^\infty \frac{(\ell-2)!}{\ell!}r^\ell \mathcal{E}_{L}(t)  n^L\,,
\ee
where~$r$ is the distance from the body,~$n^i \equiv x^i/r$ is the unit normal vector to the sphere, and~$\mathcal{E}_{L}$ denotes the symmetric trace-free multipole moments. (We adopt the multi-index notation, with~$L$ collectively denoting~$\ell$ spatial indices, such that~$\mathcal{E}_L =\mathcal{E}_{(i_1\cdots i_\ell)}$, and~$n^L = n^{i_1}\cdots n^{i_\ell}$.)

The external field tidally deforms the object, thereby inducing internal multipole moments of the form 
\be
I_{L} = \int {\mathrm d}^3x~\delta\rho(\vec x,t) r^\ell  n^L \,,
\ee
where~$\delta\rho$ is the body's mass density perturbation. Both external field and induced response can be equivalently expanded in terms of spherical harmonics, by introducing multipole moments~$\mathcal{E}_{\ell m}$ and~$I_{\ell m}$. For instance,~$\mathcal{E}_{\ell m} = \mathcal{E}_{L} \int {\mathrm d}\Omega~n^L Y^{*}_{\ell m}(\theta,\varphi)$. The total gravitational potential of the system can then be written as~\cite{PoissonWill}
\be 
\label{potential}
U_\text{\tiny tot}=
-\frac{M}{r}
-\sum_{\ell, m} Y_{\ell m} 
\left[\frac{(\ell-2)!}{\ell !} \mathcal{E}_{\ell m}  r^\ell -
\frac{(2\ell-1)!!}{\ell !} 
\frac{I_{\ell m}}{r^{\ell+1}}
 \right]\,.
\ee

Assuming adiabatic and weak external tidal forces, linear response theory implies that the response multipole moments must be proportional to the perturbed tidal moments. After performing a Fourier transform in time, the relation is
\be
\label{IkENew}
I_{\ell m}\left(\omega\right)
=-\frac{\left(\ell-2\right)!}{(2\ell-1)!!}k_{\ell m}(\omega) r_+^{2\ell+1} \mathcal{E}_{\ell m}\left(\omega\right)\,,
\ee
where~$r_+$ is the object size, and~$\omega$ is the perturbation frequency.   
The dimensionless proportionality coefficients,~$k_{\ell m}$, quantify the body's tidal response.
They admit the small-frequency expansion
\be
\label{eq:klm}
	k_{\ell m} \simeq \kappa_{\ell m} + {\mathrm i}\nu_{\ell m}\big(\omega - m \Omega\big) + \dots\,,
\ee
where~$\Omega$ the body's angular velocity. The real part represents the conservative, static response, with coefficients~$\kappa_{\ell m}$ known as TLNs. The imaginary part~$\nu_{\ell m}$ accounts for dissipative effects. In the following, we will focus on the static response of compact objects, and summarize the main predictions for Schwarzschild black holes.

\subsection{Schwarzschild black holes in $D$ dimensions}
\label{sec:sch}
\noindent
We now extend the above results to full GR, focusing on the example case of Schwarzschild BHs in~$D$ space-time dimensions, following~\cite{Hui:2020xxx}. The metric is given by
\begin{align}\label{eq:sch}
    \mathrm{d}s^2 = -f(r) \mathrm{d}t^2 + \frac{\mathrm{d}r^2}{f(r)}+r^2 \mathrm{d}\Omega_{D-2}^2\,,
\end{align}
where ${\rm d}\Omega_{D-2}^2$ is the line element on the~$D-2$ unit sphere, and 
\begin{align}
    f(r) = 1-\left(\frac{r_+}{r}\right)^{D-3}\,; \qquad r_+ = 2M\,.
\end{align}

\begin{figure}[t!]
	\centering
        \includegraphics[width=0.49\textwidth]{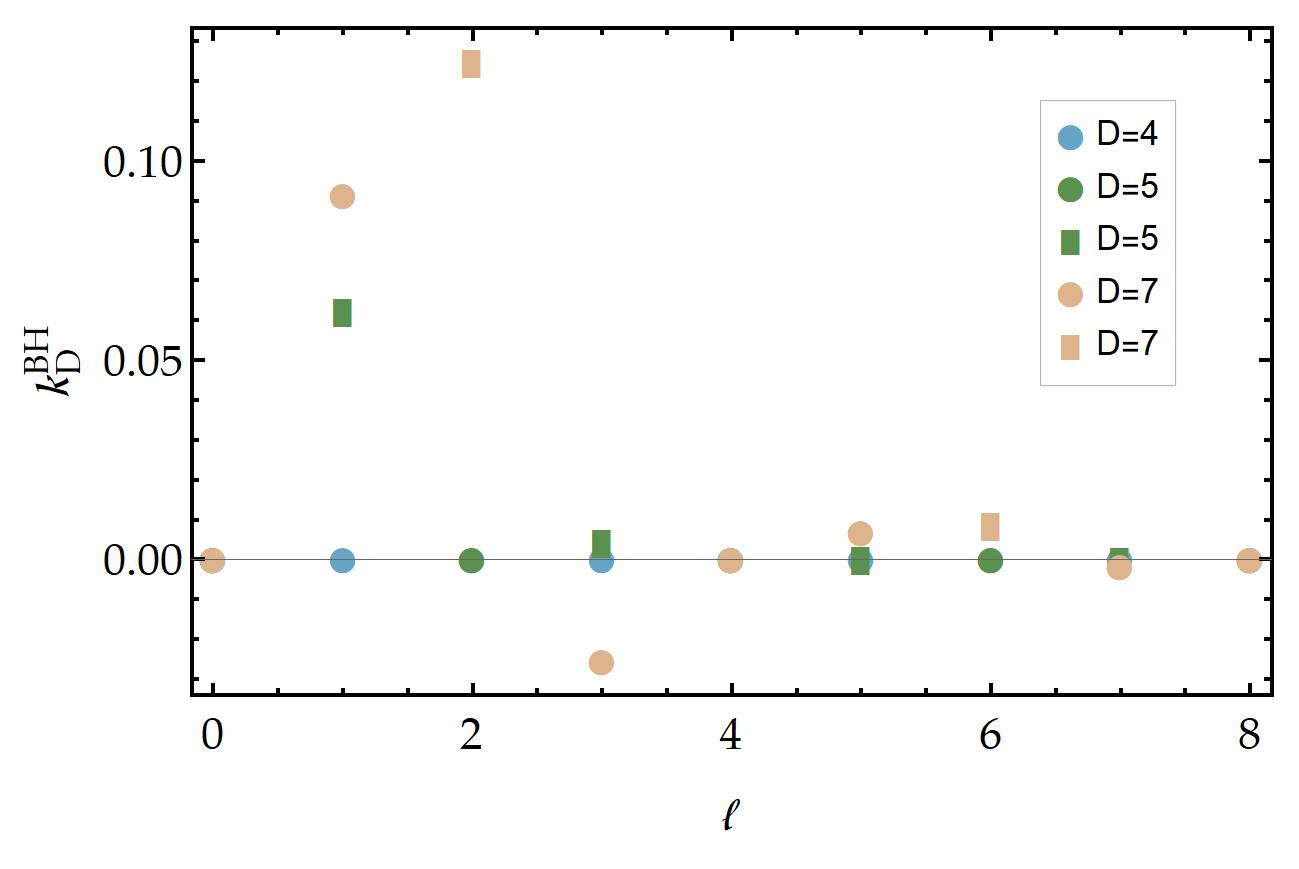}
	\caption{\it Tidal Love numbers of $D$ dimensional Schwarzschild BHs. We have factored out the logarithmic dependence $\log(r/r_0)$ for the Love numbers for those points which are plotted as rectangles; those with circles have no such running. }
	\label{fig: SchwD}
\end{figure}

The computation of BH TLNs~\cite{Deruelle:1984hq,Binnington:2009bb, Damour:2009vw} requires the analysis of linearized metric perturbations of the BH geometry~\cite{Regge:1957td,Zerilli:1970se,Teukolsky:1973ha,Moncrief:1974am,Cunningham:1978zfa,Cunningham:1979px}.
For simplicity, we focus on the proxy problem of a massless free scalar field, propagating on the fixed BH metric. The scalar wave equation is given by
\begin{align}\label{eq:waveeqn}
\partial_\mu\left(\sqrt{-g}g^{\mu\nu}\partial_\nu \phi\right) = 0\,.
\end{align}
Since we are interested in the static tidal response, the scalar can be assumed time-independent. 
Once again the spherical symmetry of the background allows us to decompose~$\phi$ in radial and angular components: 
\begin{align}
\label{phidecom}
    \phi(\vec{x}) = \sum_{\ell,m} R_{\ell m}(r) r^\frac{2-D}{2} Y_{\ell m}(\theta, \varphi)\,.
\end{align}
Substituting into~\eqref{eq:waveeqn}, the equation of motion for the radial mode function becomes~\cite{Hui:2020xxx}
\begin{align}
    &R_{\ell m}''(r) + \frac{f'(r)}{f(r)} R_{\ell m}'(r) - \left[\frac{\ell(\ell+D-3)}{f(r)\,r^2} + \frac{f'(r)}{f(r)} \frac{D-2}{2r}\right. \nonumber\\
    &\left.~~~~~~~~~~~~~~~~~~~~ +\,\frac{(D-2)(D-4)}{4r^2}\right]R_{\ell m} (r) = 0\,,
\end{align}
where primes denote radial derivatives. As shown in greater detail in the following sections, such equations can be cast in a hypergeometric form by performing the field redefinition $u_{\ell m} = (r/r_+)^{\frac{D+2\ell-4}{2(D-3)}}R_{\ell m}$. After imposing regularity at the BH horizon, the solution at large distances takes the general form 
\begin{align}
R_{\ell m} (r \to \infty) & \simeq  C \left( \frac{r}{r_+} \right)^{\frac{D}{2}-1}  \left[ \left( \frac{r}{r_+} \right)^{\ell} + \dots \right. \nonumber \\
& \left.~~~~~~~~~~~~ +~ k^\text{\tiny BH}_{\ell m} \left( \frac{r_+}{r} \right)^{\ell+D-3} + \dots \right]\,.
\end{align}
The first term in brackets, which grows as~$r^{\ell}$, is interpreted as the applied scalar field profile with overall amplitude~$C$.\footnote{In light of Eq.~\eqref{phidecom}, the quantity in square brackets gives the scaling behavior of~$\phi$.} This is the analogue of the external tidal field in the gravitational case. The second term, which falls off as~$r^{-\ell - D + 3}$, encodes the response of the BH.
As in the Newtonian case, the static TLNs~$k^\text{\tiny BH}_{\ell m}$ are defined as the ratio of the response term over the external source. Explicitly, their value depends on an effective multipole moment~$\hat{\ell} \equiv \frac{\ell}{D-3}$~\cite{Hui:2020xxx}:
\begin{equation}
\label{kBH D} k^\text{\tiny BH}_{\ell} =
\begin{cases}
 \frac{2\hat{\ell}+1}{2\pi} \frac{\Gamma (\hat{\ell}+1)^4}{\Gamma (2\hat{\ell} + 2)^2} \tan (\pi \hat{\ell}) \qquad \quad \,\,\,\,\, \text{for generic} \,\, \hat{\ell}  \\[7pt]
 \frac{(-1)^{2\hat{\ell}} (D-3) \Gamma (\hat{\ell}+1)^2}{(2\hat{\ell})! (2\hat{\ell}+1)! \Gamma(-\hat{\ell})^2} \log \left( \frac{r_0}{r} \right) \,\,\,\,\,\,\,\,\,\, \text{for half-integer} \,\, \hat{\ell}  \\[7pt]
 0 \qquad \qquad \qquad  \qquad \qquad \qquad \,\,\,\, \text{for integer} \,\, \hat{\ell} \,.
\end{cases}
\end{equation}
In particular, in~$D = 4$, corresponding to integer~$\hat{\ell}$, one recovers the known vanishing of the BH TLNs. In higher dimensions, the static TLNs vanish only for integer multiples of~$D-3$. Furthermore, it is interesting to notice the appearance of a logarithmic term in the response for half-integer~$\hat{\ell}$. This is an example of classical renormalization group running in terms of an arbitrary reference scale~$r_0$~\cite{Kol:2011vg}. The result is summarized in Fig.~\ref{fig: SchwD}. In the next sections we will use this result to compare the GR predictions to those of analogue gravity models.

\section{Basics of analogue gravity}
\label{sec: ABH}
\noindent
Analogue gravity models describe the correspondence between BHs in GR and supersonic acoustic flows, assuming the fluid to be barotropic, inviscid and locally irrotational~\cite{Visser:1997ux,Barcelo:2005fc}. In this section we summarize the main assumptions behind the derivation of acoustic metrics and highlight the connection between gravity and hydrodynamics more concretely.

As a starting point, consider the hydrodynamical equations of non-relativistic fluid mechanics~\cite{landau1974fluid,shapiro1953compressible, chandrasekhar1981hydrodynamic}. For an inviscid fluid, they are given by the familiar continuity and Euler equations,
\begin{align}
&\partial_t \rho + \vec\nabla \cdot (\rho \vec v) = 0\,; \nonumber \\
& \rho \Big(\partial_t
{\vec v}+\big({\vec v} \cdot \vec\nabla\big){\vec  v}\Big)= -\vec\nabla p+\vec{F}_{\rm ext}\,,
\end{align}
for the fluid density~$\rho$, pressure~$p$, and velocity field~$\vec{v}$. The external force~$\vec{F}_{\rm ext}$ is assumed
conservative,
\be
\vec{F}_{\rm ext} = -\rho\vec\nabla \psi_\text{\tiny ext} \,,
\ee
with the potential~$\psi_\text{\tiny ext}$ taken to be fixed and unperturbed by the fluid. Furthermore, the fluid is taken to be locally irrotational,
which allows to introduce the velocity potential~$\Phi$ as 
\be
{\vec v}=-\vec \nabla \Phi\,.
\ee
The continuity equation thus reduces to
\be
\partial_t \rho - \vec\nabla \cdot \big(\rho \vec \nabla \Phi\big) = 0 \,.
\label{cont}
\ee
Lastly, the fluid is assumed barotropic, {\it i.e.} $\rho = \rho (p)$, such that one can define the enthalpy
\be
h(p)=\int_0^p \frac{{\rm d}p'}{\rho(p')}\,.
\label{enthalpy}
\ee
These simplifying assumptions imply that the Euler equations reduce to a single scalar equation:
\be
- \partial _t \Phi +h +\frac{1}{2}\big(\vec\nabla \Phi\big)^2+\psi_\text{\tiny ext}=0\,.
\label{Euler}
\ee

Sound waves are defined as small fluid perturbations over a certain background flow. They can be studied by linearizing the fluid equations, setting $\rho = \rho_0 + \rho_1$, $p = p_0 + p_1$ and $\Phi = \Phi_0 + \phi$. At linear order, Eqs.~\eqref{cont} and~\eqref{Euler} give\footnote{Backreaction effects on the external driving potential~$\psi_\text{\tiny ext}$ are not considered, such that the latter is fixed in the perturbative expansion~\cite{Visser:1997ux}.}
\begin{align}
&\partial _t \rho _1 + \vec \nabla \cdot \left(\rho _1 \vec v_0 - \rho _0\vec\nabla \phi\right)=0\,; \nonumber \\ 
& \frac{\rho_0}{c_s^2} \left(\partial _t \phi +  \vec v_0 \cdot \vec\nabla \phi\right) = \rho_1 \,. 
\label{pert fluid eoms}
\end{align}
To obtain this, we have substituted the perturbed enthalpy,~$h \simeq h_0 + p_1/\rho_0$, and expressed~$p_1$ in terms of the adiabatic sound speed:
\be
c_s^2 = \left.\frac{{\rm d}p}{{\rm d}\rho}\right\vert_0 = \frac{p_1}{\rho_1}\,.
\ee

As usual, the linearized fluid equations can be combined into a single, second-order PDE describing the propagation of sound waves,
by using the second of Eqs.~\eqref{pert fluid eoms} to eliminate~$\rho_1$ from the first. The result is
\begin{align}
& -\partial_t \left ( \frac{\rho_0}{c_s^2}
\big(\partial_t \phi+{\vec v_0} \cdot \vec\nabla \phi\big)\right ) \nonumber \\
& + \vec\nabla \cdot \left(\rho _0 \vec\nabla \phi- \frac{\rho_0 {\vec v_0}}{c_s^2}\big(\partial_t \phi +{\vec v_0} \cdot \vec\nabla \phi\big)
\right ) = 0\,. 
\end{align}
The key insight of analogue gravity is that the above equation for the perturbed velocity potential~$\phi$ can be interpreted as 
a wave equation~\eqref{eq:waveeqn} for a scalar field on an effective Lorentzian acoustic metric~\cite{Visser:1997ux, Barcelo:2005fc}
\begin{eqnarray}
g_{\mu \nu}\equiv \frac{\rho _0}{c_s} \left[
\begin{array}{ccc}
-\big(c_s^2-v_0^2\big) &\vdots&-v_0^{j}\\
\hdots \hdots\hdots &.&\hdots \hdots\\
-v_0^{i}&\vdots&\delta _{ij}
\end{array}
\right]\,.
\label{metricvisser}
\end{eqnarray}
The associated Lorentzian line element is
\be
{\rm d}s^2 = \frac{\rho _0}{c_s} \left[ -\big(c_s^2-v_0^2\big){\rm d}t^2 + \delta_{ij} \big({\rm d}x^i - v_0^i {\rm d}t\big)\big({\rm d}x^j - v_0^j {\rm d}t\big)\right]\,.
\label{acoustic metric}
\ee
The effective geometry depends on the background fluid quantities,~$\rho_0$ and~$\vec{v}_0 = - \vec{\nabla}\Phi_0$, which satisfy the unperturbed equations:
\begin{align}
\partial_t \rho_0 - \vec\nabla \cdot \big(\rho_0 \vec \nabla \Phi_0\big) & = 0\,; \nonumber \\
- \partial _t \Phi_0 +h_0 +\frac{1}{2}\big(\vec\nabla \Phi_0\big)^2+\psi_\text{\tiny ext} & =0 \,.
\end{align}
The upshot is that propagation of sound waves in a fluid can be equivalently described by the propagation of a scalar field in a generic Lorentzian space-time given by the acoustic metric. Notice, however, that the acoustic metric is more restricted. Whereas a general Lorentzian metric in~$D$ dimensions has~$\frac{D(D-1)}{2}$ algebraically-independent components at every space-time point, the acoustic metric is parametrized by only 3, namely~$\rho_0$,~$\Phi_0$ and~$c_s$. 
This number is further reduced to 2 degrees of freedom by the background continuity equation.

From~\eqref{acoustic metric}, any region of supersonic flow (where~$v_0 > c_s$) is an ergo-region, whose boundary ($v_0 = c_s$) is an ergo-surface. Furthermore, as in GR, the boundary of the region from which null geodesics (phonons) cannot escape defines the event horizon. As usual, the event horizon coincides with the ergosphere for static metrics. 
Such a description allows for the study of phenomena analogous to those of standard BHs, such as quasinormal modes, superradiance and tail effects. See~\cite{Berti:2004ju, Cardoso:2004fi, Cardoso:2005ij, Lepe:2004kv, Kim:2004sf,Saavedra:2005ug,Abdalla:2007dz}. 

In the next sections we will extend these studies by considering the TLNs associated to various examples of ABHs. Those can be interpreted as tidal perturbations of the acoustic flow induced by the presence of an external source.

\section{(2+1) Acoustic black holes}
\label{sec: 2+1}
\noindent
We begin by investigating the hydrodynamical analogue of gravity in~$2+1$ dimensions, considering the example of a “draining bathtub” model for a rotating ABH~\cite{Visser:1997ux}. This model assumes constant~$\rho_0$, and therefore constant~$p_0$ and~$c_s$,
while the fluid velocity take the form
\begin{align}
    \vec v_0 = \frac{-A \hat{\vec r} + B\hat{\vec\varphi}}{r}\,,
\end{align}
where the constants~$A,B > 0$ parametrize the radial and angular components of the velocity field.
 The corresponding velocity potential is
\be
\Phi_0(r,\varphi) = A \ln \frac{r}{L} - B \varphi\,,
\ee
where~$L$ is an arbitrary scale.
Dropping an overall constant prefactor, the acoustic metric~\eqref{acoustic metric} becomes~\cite{Visser:1997ux}
\begin{align}
\nonumber
{\rm d}s^2 = & - \left(c_s^2 - \frac{A^2+B^2}{r^2}\right) {\rm d}t^2 + \frac{2A}{r} {\rm d}r{\rm d}t + {\rm d}r^2  \\
& - 2B {\rm d}\varphi {\rm d}t + r^2 {\rm d}\varphi^2\,. 
\label{tub 1}
\end{align}
The ergo-surface (ergo-circle) lies at~$r_\text{\tiny ergo} = \frac{\sqrt{A^2 + B^2}}{c_s}$. The ABH event horizon corresponds to the surface where the radial velocity becomes supersonic, and hence lies at\footnote{The sign choice~$A > 0$ ensures that~$r_+$ is a future horizon, whereas~$A < 0$ would imply a past horizon (acoustic white hole).}
\be
r_+ = \frac{A}{c_s}\,.
\ee
It is useful to perform the coordinate redefinition~\cite{Basak:2002aw,Berti:2004ju, Cardoso:2004fi} 
\begin{align}
{\rm d}t = {\rm d}\tilde{t} + \frac{Ar}{r^2c_s^2-A^2} {\rm d}r\,;~~~ {\rm d} \varphi = {\rm d}\tilde{\varphi} + \frac{BA}{r\big(r^2c_s^2-A^2\big)}{\rm d}r\,,
\end{align}
such that~\eqref{tub 1} takes the Kerr-like form
\begin{align}
    \mathrm{d}s^2 = &-\left[1-\frac{r_+^2}{r^2}\left(1 + \frac{B^2}{A^2}\right)\right] c_s^2 \mathrm{d}\tilde{t}^2 + \left(1-\frac{r_+^2}{r^2}\right)^{-1} \mathrm{d}r^2 \nonumber\\
    &- 2B \mathrm{d}\tilde{\varphi} \mathrm{d}t + r^2 \mathrm{d}\tilde{\varphi}^2\,.
\label{eq:2+1metric}
\end{align}

Consider now velocity perturbations of the background fluid,~$\vec{v}_1 = - \vec\nabla \phi$, induced by an external source.
The perturbed velocity potential satisfies the wave equation~\eqref{eq:waveeqn}, with acoustic metric~$g_{\mu\nu}$ given by~\eqref{eq:2+1metric}.
Given the background rotational symmetry, it is convenient to perform the multipole decomposition:
\be
\phi = \sum_m R_m(r) {\rm e}^{\mathrm{i}\left(m\tilde{\varphi}-\omega \tilde{t}\right)}\,.
\ee
From here on, we consider the static limit $\omega = 0$. Substitution into~\eqref{eq:waveeqn} yields an equation for~$R_m(r)$:
\begin{align}\label{eq:2+1_radial}
    \Delta(r) \Big(\Delta(r) R_m'(r)\Big)' + \left(\frac{B^2}{c_s^2 r^2}-\frac{\Delta(r) }{r}\right)m^2 R_m(r)=0\,,
\end{align}
where~$\Delta(r) = r -r_+^2/r$. The radial function~$R_m(r)$ carries the information of the external source and ABH response, allowing us to extract the associated TLNs, following the computation in Sec.~\ref{sec: TLN-GR}.

With a change of variable,~$z=\frac{r_+^2}{r^2}$, and a redefinition  
\be
R_m(z) = z^{\alpha}(1-z)^{\beta} F(z) \,,
\label{RF redef}
\ee
with~$\alpha = |m|/2$ and~$\beta = \frac{\mathrm{i} B |m|}{2 r_+c_s}$, Eq.~\eqref{eq:2+1_radial} is recast in the 
hypergeometric form~\cite{abramowitz+stegun} (where~$' = {\rm d}/{\rm d}z$)
\begin{align}
\label{eq:hypergeometric}
    z(1-z)F''(z)+ \big(c-(1+a+b)z\big)F'(z){{\rm d}z} - abF(z)=0\,,
\end{align}
with coefficients
\begin{align}
    a=1+\frac{|m|}{2}+\frac{\mathrm{i} B |m|}{2r_+ c_s}\,; \quad b=a-1\,; \quad c=1+|m|\,.
\label{abc}
\end{align}
As usual, this admits two independent solutions. However, one of these is eliminated by requiring that the radial wavefunction does not diverge at the horizon $z = 1$, and thus satisfies the ingoing boundary condition associated to BH perturbation theory~\cite{Berti:2009kk}. The regular solution then reads
 \begin{align}
 \label{sol2+1}
    R_m(z) = C z^{\frac{|m|}{2}}(1-z)^{\frac{\mathrm{i} B |m|}{2 r_+c_s}}  \times {}_2F_1(a,b;\tilde{c};1-z)\,,
\end{align} 
where~$\tilde{c} = 1+a+b-c$, and~$C$ is an overall constant. 

The study of the asymptotic $z \to 0$ ($r\rightarrow \infty$) behavior of the solution carries the necessary information for the computation of the TLNs. One must, however, proceed with caution and distinguish the special cases where $a,b,c$, or some combination thereof, are integers. From Eq.~\eqref{abc}, $c$ is always an integer, and therefore so is~$\tilde{c} - a - b$, while $a,b$ are integers only for non-rotating ABHs~($B = 0$) and even~$m$. Since~$\tilde{c}-a-b$ is an integer, 
the asymptotic expression of Eq.~\eqref{sol2+1} reads~\cite{abramowitz+stegun}
\begin{align}
     R_m(z \to 0)  & \simeq C z^{\frac{|m|}{2}} \times \bigg[ {}_2F_1 (a,b,c;z) \log z  \nonumber \\
    & + \frac{(-1)^{|m|+1} (c-1)!(c-2)!}{(1-a)_{|m|}(1-b)_{|m|}}z^{-|m|} \nonumber \\
   & + \psi(a)+\psi(b)-\psi(1)-\psi(c) \bigg]\,,
\end{align}
where $(d)_n$ denotes the Pochhammer symbol, and $\psi$ is the digamma function. The asymptotic expansion allows us to identify the growing term, proportional to $z^{-|m|}$, and the decaying term, proportional to $\log z$, so that the relative coefficient gives the TLN\footnote{Note that the same running in the TLN would appear by performing analytic continuation, {\it i.e.}, generalizing the multipole number~$m$ to be real, assuming the standard expansion of hypergeometric functions through the Kummer property of Eq.~\eqref{Kummer}, and then performing a renormalization scheme procedure similar to the one of Refs.~\cite{Kol:2011vg, Rodriguez:2023xjd}.}
\begin{widetext}
\begin{equation}
k^\text{\tiny (2+1)\,ABH}_m = \frac{2 (-1)^{-|m|} \Gamma \left[\frac{|m|}{2} \left(1-{\rm i}\tilde{B}\right)\right] \Gamma \Big[1+\frac{|m|}{2}\left(1- i \tilde B\right)\Big]}{\Gamma \big(|m|\big) \Gamma \big(|m|+1\big) \Gamma \left[1-\frac{|m|}{2} \left(1+{\rm i} \tilde{B}\right)\right] \Gamma \left[-\frac{|m|}{2} \left(1+{\rm i} \tilde{B}\right)\right]} \log \left(\frac{r}{r_0}\right)\,; \qquad \tilde{B} \equiv \frac{B}{c_s r_+} \,,
\label{k2+1}
\end{equation}
\end{widetext}
where~$r_0$ is an arbitrary reference scale. It is interesting to notice that the TLN displays a logarithmic running  with radial distance, similar to Schwarzschild BHs in~$D$ dimensions for half-integer $\hat{\ell}$ (see the middle line in Eq.~\eqref{kBH D}). Notice that the digamma functions could be absorbed into~$r_0$. Furthermore, in the non-rotating case $B = 0$, one obtains vanishing TLNs for even values of $m$, as expected from the properties of the hypergeometric solutions when~$a$ and~$b$ are integers. 

The results are summarized in the left panel of Fig.~\ref{fig:3+1}, for different values of~$\tilde{B}$. At fixed~$|m|$, we see that the TLNs grow monotonically with~$\tilde{B}$. Notice also that only for sufficiently large values of~$\tilde{B}$ do the TLNs grow with increasing~$|m|$. Furthermore, it is instructive to compare these results to the prediction of a Schwarzschild BH in $D=5$ space-time dimensions, as reviewed in Sec.~\ref{sec:sch}. This comparison is interesting because in the nonspinning case $B=0$, the radial dependence of the metric function $f(r) = 1 - r_+^2/r^2$ is the same in both cases, although the angular line element for the Schwarzschild BH lies on a $3$-sphere. The evolution of the ABH TLNs across multipole moments (Fig.~\ref{fig:3+1}, left panel, yellow points) is therefore similar to the one of standard BHs, as shown by the green points in Fig.~\ref{fig: SchwD}, although not identical. A similar comparison would hold with non-rotating black strings in~$D = 6$ dimensions~\cite{Rodriguez:2023xjd}.

\section{(3+1) Acoustic black holes}
\label{sec: 3+1}
\noindent
We now generalize the discussion of ABHs to 3+1 space-time dimensions, in order to provide a more direct comparison to GR BHs. Given that ABHs are unable to perfectly replicate the dynamics of BHs, since ABHs are constrained by the fluid equations instead of Einstein's equations, we expect a different prediction for the corresponding TLNs, which we compute in this section. Nevertheless, the TLNs of ABHs do share some similarities with those of BHs, such as vanishing for certain multipoles, which can be explained via ladder symmetries. We will consider the most general case of canonical ABHs, and then discuss the simplified de Laval nozzle model.

\subsection{Canonical ABH}
\label{sec: 3+1 canonical}
\noindent
The most general spherically-symmetric acoustic metric describing the flow of an incompressible fluid is given by~\cite{Visser:1997ux}
\begin{align}\label{eq:canonical_metric}
    \mathrm{d}s^2 = - f(r) c_s^2  \mathrm{d}t^2 + \frac{\mathrm{d}r^2}{f(r)}+r^2 \mathrm{d}\Omega_2\,,
\end{align}
with
\begin{align}
f(r) = 1-\left(\frac{r_+}{r}\right)^4\,,
\label{canonicalf(R)}
\end{align}
where the horizon scale~$r_+$ is a constant. This assumes constant density~$\rho_0$ and sound speed~$c_s$, and a background velocity profile of the form
\begin{align}
v_0 = c_s \left(\frac{r_+}{r}\right)^2\,.
\end{align}

Similarly to the previous section, we focus on time-independent perturbations and perform the spherical harmonic decomposition~$\phi(\vec x) = \sum_{\ell,m} R_{\ell m} (r) Y_{\ell m}(\theta,\varphi)$. The radial function then satisfies the equation of motion
\begin{align}
\label{eq:canonical_radial}
    \frac{1}{r^2}\Big(r^2 f(r)R_{\ell m}'(r)\Big)'- \frac{ \ell(\ell+1)}{r^2} R_{\ell m}(r)=0\,.
\end{align}
Introducing the variable $z=1-(r_+/r)^4$, and the redefinition~\eqref{RF redef} with~$\alpha = 0$ and~$\beta = -\ell/4$,
this can be recast into the hypergeometric form~\eqref{eq:hypergeometric} with
\begin{align}
\label{abccan}
    a=\frac{3-\ell}{4}\,; \qquad b=-\frac{\ell}{4}\,; \qquad c = 1\,.
\end{align}
The solution satisfying ingoing boundary conditions at the BH horizon~$z = 0$ is given by
\begin{equation}
\label{Rcan}
R(z) = C (1-z)^{-\ell/4}{}_2 F_1(a,b,1;z)\,,
\end{equation}
where~$C$ is again a constant.

The TLNs can be read off as before from the asymptotic behavior~$z \to 1$ ($r\rightarrow \infty$). Since~$c-a-b = \frac{1+2\ell}{4}$ is non-integer,
we can use the known Kummer property~\cite{abramowitz+stegun},
\begin{widetext}
    \begin{align}
    \label{Kummer}
{}_2 F_1(a,b,c;z) & =  \frac{\Gamma(c)\Gamma(c-a-b)}{\Gamma(c-a)\Gamma(c-b)}{}_2 F_1(a,b,a+b-c+1;1-z) \nonumber \\
        & + (1-z)^{c-a-b}\frac{\Gamma(c)\Gamma(a+b-c)}{\Gamma(a)\Gamma(b)}{}_2 F_1(c-a,c-b,c-a-b+1;1-z)\,.
    \end{align}
\end{widetext}
One can then easily separate the growing and decaying modes, and determine the TLNs as
\begin{align}
\label{TLNABHcan}
    k^\text{\tiny (3+1)\,Canon.}_{\ell} = \frac{\Gamma \left(-\frac{\ell}{2}-\frac{1}{4}\right) \Gamma \left(\frac{\ell}{4}+1\right) \Gamma \left(\frac{\ell+1}{4}\right)}{\Gamma \left(\frac{3}{4}-\frac{\ell}{4}\right) \Gamma \left(\frac{\ell}{2}+\frac{1}{4}\right) \Gamma \left(-\frac{\ell}{4}\right)}\,.
\end{align}
As expected, the TLNs vanish for values of~$\ell$ such that~$a,b$ are integers, {\it i.e.}, in this case~$\ell = 4 n$ or $\ell = 3+4n$ with $n \in \mathbb{N}$, while they are non-zero in general. 

\begin{figure*}[t!]
	\centering
        \includegraphics[width=0.49\textwidth]{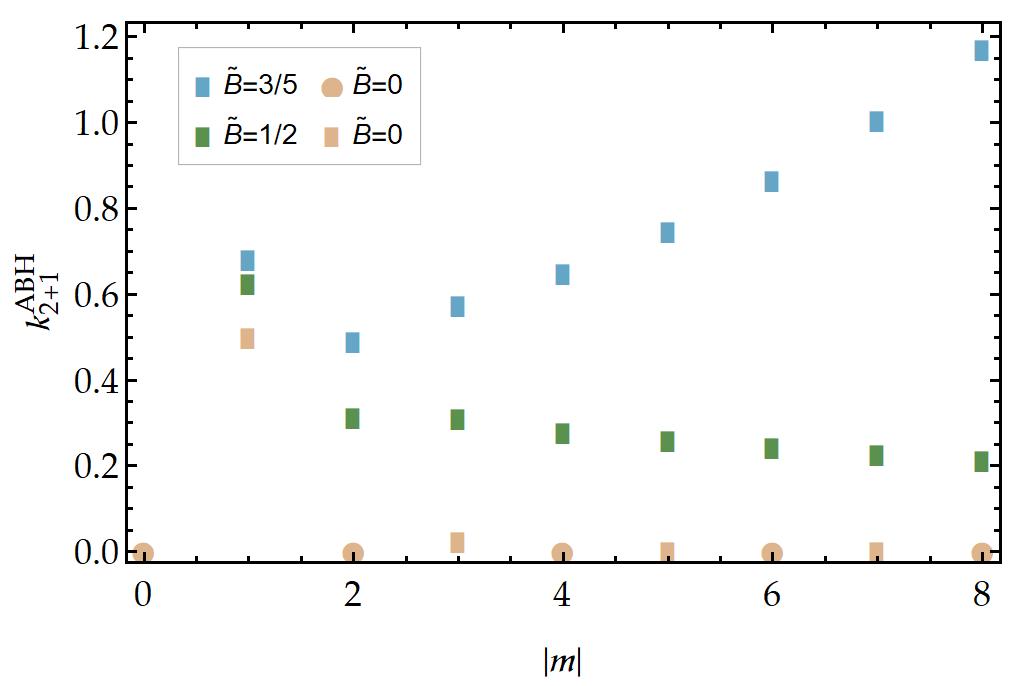}
 	\includegraphics[width=0.49\textwidth]{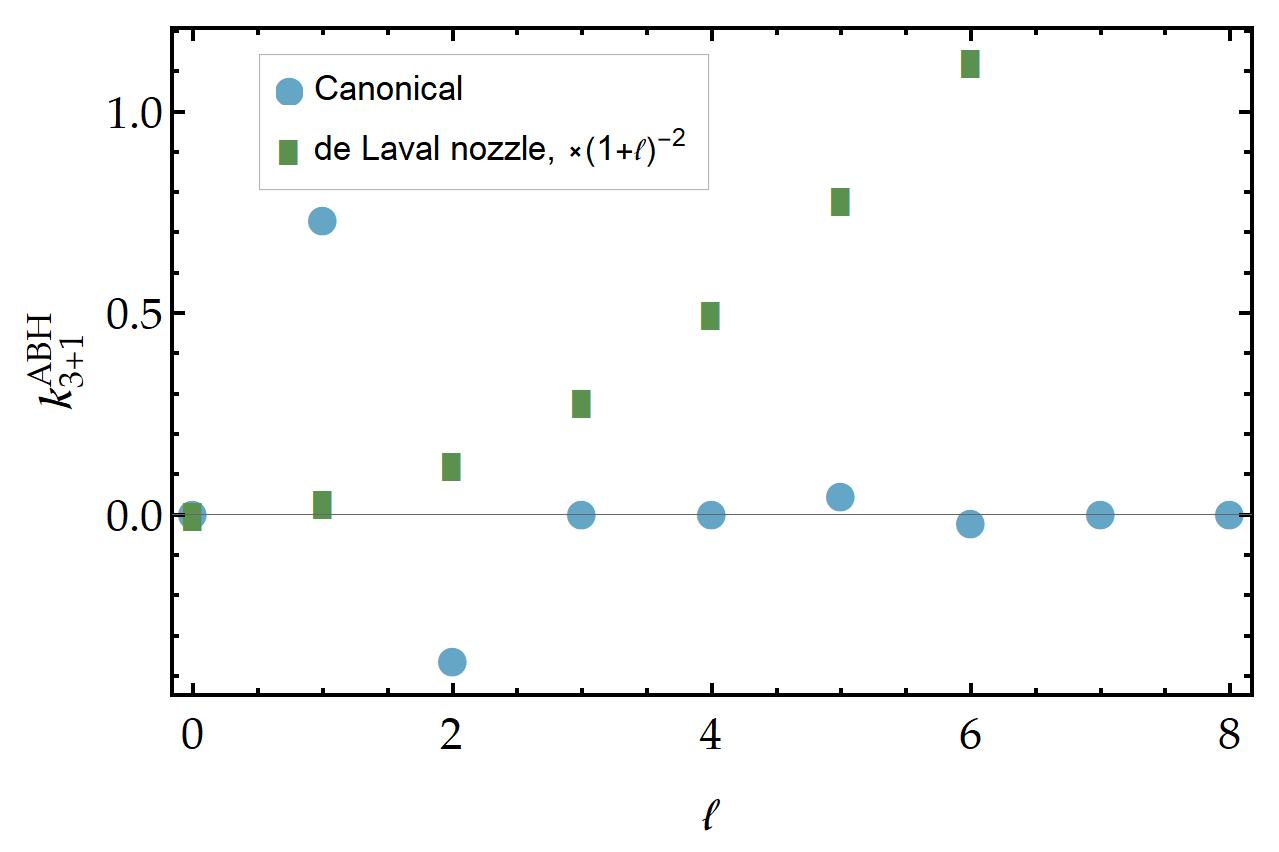}
	\caption{\it Tidal Love numbers of 2+1 ABHs (left), where~$\tilde{B}$ is the spinning parameter (Eq.~\eqref{k2+1}), and 3+1 ABHs (right). Similarly to Fig.~\ref{fig: SchwD}, we have factored out the logarithmic dependence for the Love numbers $\log(r/r_0)$ for the rectangular points. The de Laval nozzle metric assumes~$\tilde{A} = 2 c_s/r_+$.}
	\label{fig:3+1}
\end{figure*}

The results are plotted in the right panel of Fig.~\ref{fig:3+1} (blue dots). It is instructive to compare these with the predictions of~$D=7$ Schwarzschild BHs (yellow points of Fig.~\ref{fig: SchwD}), whose metric function~$f(r)= 1 - \left(\frac{r_+}{r}\right)^4$ coincides with Eq.~\eqref{canonicalf(R)}. While the results are different in detail, owing to the different dimensionality of the sphere, there are important similarities. In particular, the TLNs of~$D = 7$ BHs also vanish for~$\ell = 4 n$, {\it i.e.}, in step size of~$\Delta\ell = 4$. A key difference is that canonical ABHs have an additional tower of vanishing TLNs, for~$\ell = 3+4n$. This structure will be understood as a consequence of ladder symmetries in Sec.~\ref{sec: Ladder}. 


\subsection{de Laval nozzle}
\label{sec: 3+1 Laval}
\noindent
A simplified version of the canonical ABH is based on the de Laval nozzle, a device used to make a smooth transition from subsonic to supersonic flow. It is described by the radial expansion of the velocity field near the ABH horizon~\cite{Kim:2004sf,Saavedra:2005ug,Abdalla:2007dz}
\begin{align}
\label{Lavalvelo}
    v_0^r = -c_s + \tilde A (r-r_+) +\mathcal{O}(r-r_+)^2\,,
\end{align}
where $\tilde A = \vec\nabla\cdot \vec v_0\big|_{r=r_+} > 0$. 
The corresponding metric is  a special case of the near-horizon expansion of the canonical ABH metric of Eq.~\eqref{eq:canonical_metric}, with the metric function
\begin{align}
\label{metricLvalal}
    f(r) = \frac{2 \tilde A}{c_s}\big(r-r_+\big)\,.
\end{align}
An equivalence with the near-horizon region of the canonical metric is obtained by setting $\tilde A = 2c_s/r_+$.
The upshot of the metric function is its obvious resemblance to the Schwarzschild BH in~$D =4$ dimensions. Although only valid in the near-horizon limit, following earlier work~\cite{Kim:2004sf,Saavedra:2005ug,Abdalla:2007dz} we will take Eq.~\eqref{metricLvalal} as defining the geometry out to large distances.\footnote{In general, this would require a conspiracy among higher-order terms in Eq.~\eqref{Lavalvelo}.}

The procedure of the previous subsection allows us to write down the radial solution of the de Laval nozzle problem similarly to Eq.~\eqref{RF redef},
with~$\alpha=\beta=0$ and~$z=1-r_+/r$. The hypergeometric equation has parameters
\begin{align}
    a,b=\frac{1}{2}\left(-1\pm \sqrt{1-\frac{2\ell(\ell+1)c_s}{\tilde A r_+}}\,\right)\,; \qquad c=1\,.
\end{align}
Except for~$\ell = 0$, we see that~$a,b$ are always non-integer, while~$c-a-b = 2$. The appropriate linear transformation property in this case is~\cite{abramowitz+stegun}
\begin{align}
\label{Kummer2}
&   {}_2 F_1 (a,b,a+b+2;z) \simeq  \frac{\Gamma(a+b+2)}{\Gamma(a+2)\Gamma(b+2)} \Big( 1 - ab(1-z)\Big) \nonumber \\
 & ~~~~~~  -\frac{\Gamma(a+b+2)}{2\Gamma(a)\Gamma(b)}(1-z)^2\bigg[\log(1-z) \nonumber \\
&~~~~~~~~~~~~~~    - \psi(1)-\psi(3)+\psi(a+2)+\psi(b+2)\bigg]\,,
\end{align}
truncated to consider only the leading decaying term~$\propto (1-z)^2$. The second term on the first line, proportional to~$1-z$,
is interpreted as a subleading correction to the source, and as such can be ignored. Hence the TLNs can be read off as 
\begin{align}
    k^\text{\tiny (3+1)\,de Laval}_{\ell} &= \frac{\Gamma(a+2)\Gamma(b+2)}{2\Gamma(a)\Gamma(b)} \log \left( \frac{r}{r_0}\right) \nonumber \\
    & = \frac{c_s^2 \ell^2 (\ell+1)^2}{8 \tilde{A}^2 r_+^2}     
    \log \left( \frac{r}{r_0}\right)\,.
\end{align}
This shows that the TLNs of de Laval nozzle ABHs exhibit logarithmic running for all $\ell$ and grow with increasing $\ell$. This is shown in the right panel of Fig.~\ref{fig:3+1}. This result recovers the vanishing of the TLN for the~$\ell = 0$ mode, as shown in Ref.~\cite{Kim:2004sf} through the reflection coefficient in the scattering of acoustic waves. 

\section{Ladder symmetries}
\label{sec: Ladder}
\noindent

We have seen that the static TLNs of ABHs vanish for special multipole values, specifically for even values of~$m$ in the case of (2+1)-dimensional nonrotating ABHs, and for~$\ell = 4n$ and~$4n+3$~($n \in \mathbb{N}$) in the case of~(3+1)-dimensional canonical ABHs. It is natural to ask whether this is a consequence of an underlying symmetry of the perturbation equations. 

In the context of asymptotically flat GR BHs, the vanishing of static TLNs has been related to a ladder symmetry~\cite{Hui:2021vcv,Berens:2022ebl,Rai:2024lho, Combaluzier-Szteinsznaider:2024sgb}. This symmetry constrains solutions of the perturbation equations to take a simple polynomial form in the radial coordinate, thus enforcing the vanishing of TLNs.

\begin{figure}[t!]
	\centering
        \includegraphics[width=0.49\textwidth]{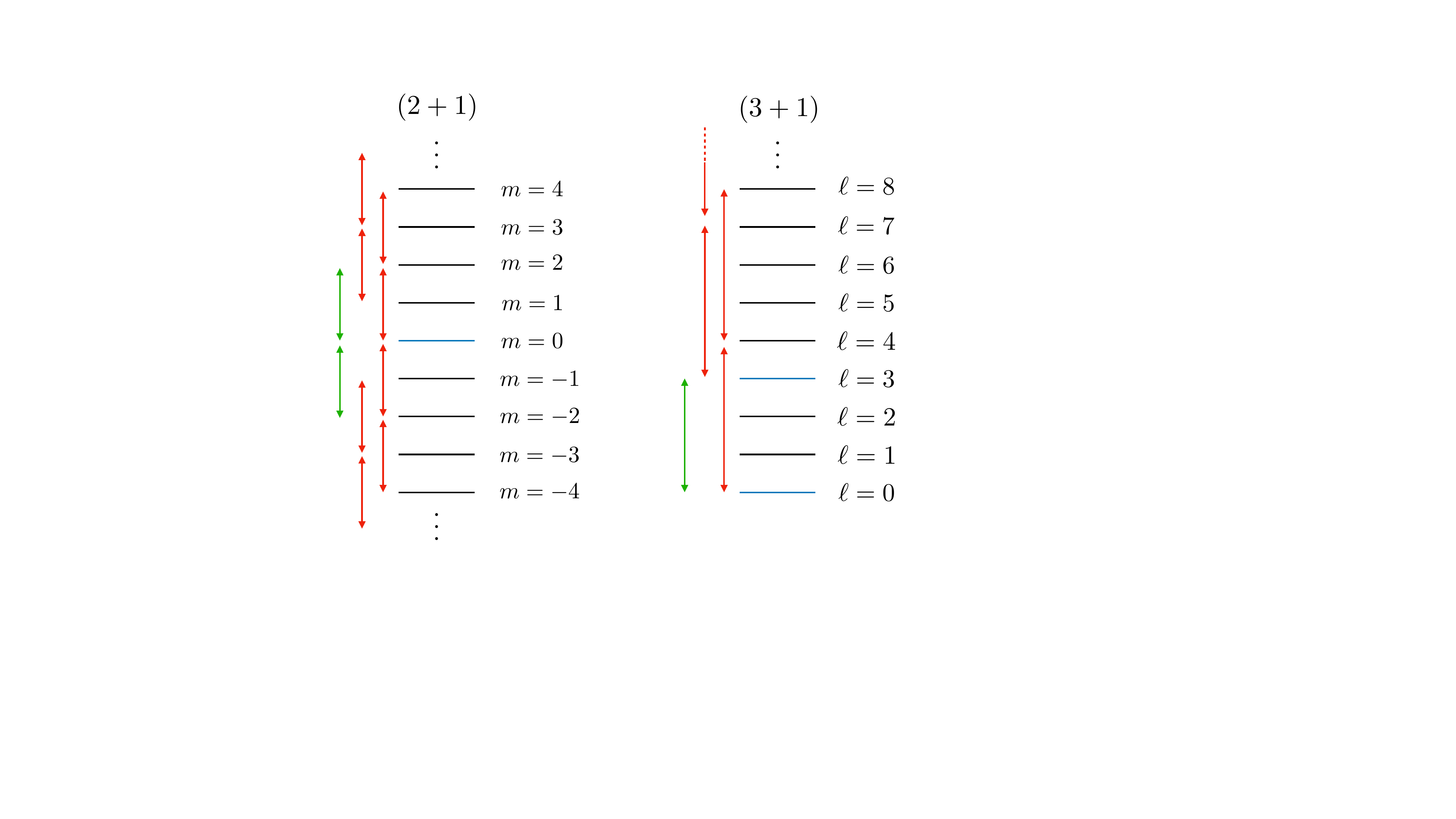}
	\caption{\it Schematic representation of the ladder structure for the (2+1) nonspinning and (3+1) canonical ABHs.
The red and green arrows denote the allowed jumps in the big and small ladders, respectively, while the blue lines show the  states with vanishing TLNs from which each ladder is built.}
	\label{fig:ladder}
\end{figure}

In this Section, we will show that a similar ladder symmetry structure is present for the (2+1) nonspinning and (3+1) canonical ABHs.
We anticipate that, contrarily to asymptotically flat GR BHs, ABHs feature two ladder structures, in analogy to the response coefficients of perturbations in theories characterized by screening mechanism, such as Born-Infeld electromagnetism or Dirac-Born-Infeld scalar theory~\cite{BeltranJimenez:2022hvs, BeltranJimenez:2024zmd}. A schematic illustration of these ladder structures is shown in Fig.~\ref{fig:ladder}.

\subsection{(2+1) Non-spinning}
\noindent
The starting point is to rewrite the perturbation equation~\eqref{eq:2+1_radial} in the form $H_m R_{m} =0$, in terms of the Hamiltonian operator
\begin{equation}
H_m \equiv - r^4 \left[\Delta (r) \frac{{\rm d}}{{\rm d}r} \left(\Delta (r) \frac{{\rm d}}{{\rm d}r}   \right)- m^2 \frac{\Delta (r)}{r} \right]\,,
\end{equation}
where we recall that~$\Delta (r) = r -r_+^2/r$. Since this depends on~$m^2$, we henceforth assume without loss of generality that~$m >0$. (The case~$m <0$ is obtained by the replacement~$m\rightarrow -m$ in all expressions below.) Let us define the ladder operators~$D^\pm_{m}$ as
\begin{align}
D^\pm_{m} \equiv \mp r^2 \Delta (r) \frac{{\rm d}}{{\rm d}r}   - m r^2 + \frac{m^2}{2( m\pm 1)} r_+^2 \,.
\end{align}
The operators~$D^\pm_m$ are well-defined for all values of~$m$, except~$m = \mp 1$. We will come back to this point below, which we believe has physical significance.
From their definition, it is easy to show that~$D^\pm_m$ satisfy the relation
\begin{equation}
D^-_{m+2} D^+_m - D^+_{m-2} D^-_m =  \frac{m^3\big(m^2-2\big)}{\big(m^2-1\big)^2}r_+^4\,.
\end{equation}
The Hamiltonian can then be recast as
 \begin{align}
 H_m & = D^-_{m+2} D^+_m - \frac{m^2(m+2)^2}{4(m+1)^2}r_+^4 \nonumber \\
 & =  D^+_{m-2} D^-_m - \frac{m^2 (m-2)^2}{4(m-1)^2}r_+^4\,.
\end{align}
The $D^\pm_m$ operators are recognized as raising and lowering operators,
since their commutation through the Hamiltonian gives
\be
\label{HD}
H_{m \pm 2} D^\pm_m = D^\pm_m H_m \,.
\ee
Thus, given a solution~$R_{m}$ at level~$m$, we can generate solutions at levels~$m\pm 2$ by applying the raising/lowering operator.
In other words,
\be
H_m R_{m} = 0 ~~ \implies~~ H_{m \pm 2} \big(D^\pm_m R_{m}\big) = 0 \,,
\ee
such that~$D^\pm_m R_{m} = C_m R_{m\pm 2}$ for some constant~$C_m$. Notice that~$D^-_{m+2} D^+_{m} R_{m}$ and~$D^+_{m-2} D^-_m R_{m}$ both reproduce the initial solution $R_{m}$, up to an overall constant. In the following, we will refer to this ladder structure as the ``big'' ladder, adopting the nomenclature of Refs.~\cite{BeltranJimenez:2022hvs, BeltranJimenez:2024zmd}.

Given the ladder structure, we can then identify the ground state solution with vanishing TLNs, such that every level obtained by acting with raising/lowering operators will have the same property. For the (2+1)-dimensional nonspinning ABH, the ground state is the level $m=0$, which admits a constant regular solution, $R_m = {\rm constant}$, so that the property of vanishing TLNs for this level can be extended to the solutions with even multipole values, $m = 2 n$ with $n \in \mathbb{Z}$, recovering the results of Sec.~\ref{sec: 2+1}.

For odd values of~$m$, we obtain two disconnected ladders: one for positive~$m$, and another for negative~$m$. This is because~$D^\pm_m$ are ill-defined for~$m = \mp 1$, as mentioned earlier, hence we cannot lower/raise the~$m= \pm 1$ solution to the level~$m = \mp 1$. Physically, this suggests that the raising/lowering operation cannot change the sign of angular momentum discontinuously ({\it i.e.}, without passing through~$m = 0$, which is only possible for even~$m$).

Before going to the (3+1) Canonical ABH, it is important to highlight the existence of a further ladder structure, which we will dub ``small'' ladder~\cite{BeltranJimenez:2022hvs, BeltranJimenez:2024zmd}. This new set of ladder operators, contrarily to the big one featuring jumps of $\Delta m = 2$, allows to connect states with multipoles $m \leq D -1 = 2$. These operators read
\begin{align}
\tilde{D}^+_m &\equiv -r^2 \Delta (r) \frac{{\rm d}}{{\rm d}r} + m r^2 +\frac{m^2}{2(1-m)} r_+^2 \,;  \nonumber \\
\tilde{D}^-_m &\equiv r^2 \Delta (r) \frac{{\rm d}}{{\rm d}r}  + (m-2) r^2 + \frac{(m-2)^2}{2(1-m)} r_+^2  \,.
\end{align}
They factorize the Hamiltonian~\eqref{Hell} as
 \begin{align}
 H_{2-m} & = \tilde{D}^+_{m} \tilde{D}^-_m - \frac{m^2(m-2)^2}{4(1-m)^2} r_+^4; \\
H_{m} & = \tilde{D}^-_{m} \tilde{D}^+_m - \frac{m^2(m-2)^2}{4(1-m)^2} r_+^4 \,,
\end{align}
and satisfy the commutation relations
\begin{align}
H_{2-m} \tilde{D}^+_m &= \tilde{D}^+_m H_m \,, \nonumber \\
H_{\ell} \tilde{D}^-_m &= \tilde{D}^-_m H_{2-m}\,, \nonumber \\
\tilde{D}^-_\ell \tilde{D}^+_m &= \tilde{D}^+_{2-m} \tilde{D}^-_{2-m}\,.
\end{align}
These ladder operators $\tilde{D}^{\pm}_m$ allow to relate the solutions with multipoles $m$ and $D-1 - m$. Notice, however, that the small ladder is ill-defined for the multipole $m = 1$ (and for its counterpart in the regime $m < 0$, $m = -1$), which is connected to itself in the small ladder as $m = 1 \to D-1 -m = 1$.
A schematic illustration of the small and big ladders for the (2+1) nonspinning ABH is shown in Fig.~\ref{fig:ladder}.

\subsection{(3+1) Canonical}
\noindent
Similarly, for (3+1)-dimensional Canonical ABHs, the radial equation~\eqref{eq:canonical_radial} can be rewritten in the Hamiltonian form $H_\ell R_{\ell m} =0$, with
\begin{equation}
H_\ell \equiv - r^6 \left[\Delta (r) \frac{{\rm d}}{{\rm d}r} \left(\Delta (r) \frac{{\rm d}}{{\rm d}r}  \right)- \ell (\ell+1) \Delta (r) \right]\,,
\label{Hell}
\end{equation}
where~$\Delta (r) = r^2 \left(1-\frac{r_+^4}{r^4}\right)$ in this case. Introducing the ladder operators
\begin{align}
D^+_\ell &\equiv -r^3 \Delta (r) \frac{{\rm d}}{{\rm d}r}   - (\ell+1) r^4 +\frac{(\ell+1)^2}{2 \ell+5} r_+^4 \,;  \nonumber \\
D^-_\ell &\equiv r^3 \Delta (r) \frac{{\rm d}}{{\rm d}r}   - \ell r^4 + \frac{\ell^2}{2 \ell-3} r_+^4  \,,
\end{align}
it is easy to show that these satisfy
\begin{align}
 & D^-_{\ell+4} D^+_\ell   - D^+_{\ell-4} D^-_\ell \nonumber \\
 &  = \frac{8 (\ell-2) (\ell+3) (2 \ell+1) \big(2 \ell (\ell+1)-3\big) }{\big(4 \ell(\ell + 1)-15\big)^2} r_+^8 \,.
\end{align}
The Hamiltonian~\eqref{Hell} can be expressed as
 \begin{align}
 H_\ell & = D^-_{\ell + 4} D^+_\ell - \frac{(\ell+1)^2 (\ell+4)^2}{(2 \ell+5)^2} r_+^8 \nonumber \\
 & =  D^+_{\ell-4} D^-_\ell - \frac{\ell^2  (\ell-3)^2 }{(2 \ell-3)^2} r_+^8 \,.
\end{align}
Once again, we recognize~$D^\pm_\ell$ as raising and lowering operators, with the commutation through the Hamiltonian given by
\be
H_{\ell \pm 4} D^\pm_\ell = D^\pm_\ell H_\ell \,.
\ee
Starting from a level-$\ell$ solution~$R_{\ell m}$, we can raise it to a level~$\ell+4$ through $D^+_\ell R_{\ell m}$, or lower it to a level $\ell-4$ solution
through $D^-_\ell R_{\ell m}$, such that~$D^\pm_\ell R_{\ell m} = \mathcal{C}_{\ell m}^\pm R_{\ell \pm 4 \, m}$, with overall constants~$\mathcal{C}_{\ell m}^\pm$.\footnote{The proportionality constants~$\mathcal{C}_{\ell m}^\pm$ should enforce various physical constraints. Since the action of~$D^-_\ell$ should only be valid for~$\ell \geq 4$, we expect~$\mathcal{C}_{\ell m}^- = 0$ for~$\ell < 4$. Similarly,~$\mathcal{C}_{\ell m}^-$ must vanish unless~$m$ satisfies~$|m| \leq \ell -4$. See a related discussion in Footnote 18 of Ref.~\cite{Hui:2021vcv}.} We will refer to this ladder structure as the ``big'' ladder.

It is straightforward to see from Eq.~\eqref{abccan} that the~$\ell = 0$ and~$\ell = 3$ solutions have only non-decaying solutions, given by~$R_{\ell = 0 \, m} (r) = {\rm constant}$, and~$R_{\ell = 3 \, m} (r) \sim (r/r_+)^3$, respectively, such that their associated static TLNs vanish. (This can be seen readily from Eq.~\eqref{TLNABHcan}.) Applying the raising operator, these give rise to two towers of~$\ell$ modes with vanishing TLNs, namely $\ell = 4n$ and~$4n +3$,~$n \in \mathbb{N}$, which recovers the results of the previous Section.\footnote{Let us stress that there is no obstruction in connecting the monopole or octupole spaces of solutions with the higher levels in the big ladder, given that $D^-_{\ell + 4}D^+_\ell R_{\ell = 0,3 \, m} \propto R_{\ell = 0,3 \, m}$. However, this feature has been shown to appear when studying the response of objects in theories with screening mechanisms~\cite{BeltranJimenez:2022hvs, BeltranJimenez:2024zmd}.}

Similarly to (2+1) ABHs, it is possible to identify the existence of a ``small'' ladder. Contrarily to the big one, which features jumps of $\Delta \ell = 4$, there is a new set of ladder operators which allows to connect the states with multipoles $\ell \leq D -1 = 3$. These operators read
\begin{align}
\tilde{D}^+_\ell &\equiv -r^3 \Delta (r) \frac{{\rm d}}{{\rm d}r} + \ell r^4 +\frac{\ell^2}{3-2 \ell} r_+^4 \,;  \nonumber \\
\tilde{D}^-_\ell &\equiv r^3 \Delta (r) \frac{{\rm d}}{{\rm d}r}  + (\ell-3) r^4 + \frac{(\ell-3)^2}{3-2 \ell} r_+^4  \,,
\end{align}
which factorize the Hamiltonian~\eqref{Hell} as
 \begin{align}
 H_{3-\ell} & = \tilde{D}^+_{\ell} \tilde{D}^-_\ell - \frac{\ell^2(\ell-3)^2}{(3-2\ell)^2} r_+^8; \\
H_{\ell} & = \tilde{D}^-_{\ell} \tilde{D}^+_\ell - \frac{\ell^2(\ell-3)^2}{(3-2\ell)^2} r_+^8 \,,
\end{align}
and satisfy the commutation relations
\begin{align}
H_{3- \ell} D^+_\ell &= \tilde{D}^+_\ell H_\ell \,, \nonumber \\
H_{\ell} \tilde{D}^-_\ell &= \tilde{D}^-_\ell H_{3-\ell}\,, \nonumber \\
\tilde{D}^-_\ell \tilde{D}^+_\ell &= \tilde{D}^+_{3-\ell} \tilde{D}^-_{3-\ell}\,.
\end{align}
The small ladder operator $\tilde{D}^+_\ell$ permits to connect the $\ell$ solution with the $D-1 - \ell$ multipole (and vice versa for $\tilde{D}^-_\ell$), so that one can relate the 
regular monopole solution $\ell = 0$ with the $\ell = 3$ multipole, thus providing a connection between the two towers of states built from these levels through the big ladder. This structure is schematically shown in Fig.~\ref{fig:ladder}.\footnote{Notice that the small ladder is ill-defined for the multipole $\ell = 3/2$, connecting it to itself $\ell = 3/2 \to D-1 - \ell = 3/2$.}

It is instructive to compare these results with asymptotically flat BHs in GR. In~$D = 4$ dimensions, the ladder structure raises~$\ell$ by one unit from the ground state $\ell = 0$, such that the static TLNs vanish for all~$\ell$. In contrast, the ladder structure of (3+1) canonical ABHs has a step size of $\Delta \ell = 4$. Interestingly, this coincides with the step size of GR BHs in~$D = 7$, which indeed has the same radial metric function~$f(r) = 1 - \left(\frac{r_+}{r}\right)^4$. A key difference is that canonical ABHs have two towers of vanishing TLNs, whereas~$D=7$ GR BHs have only one, and that there is no small ladder for GR BHs.

\section{Conclusions}
\label{conclusions}
\noindent
Analogue gravity models are based on the map between various aspects of the physics of GR BHs and the theory of supersonic acoustic flows. This correspondence is motivated by the appearance of an event horizon for sound waves, which leads to wave-like phenomena such as quasinormal modes, superradiance, Hawking radiation, and tail effects. Despite the strong similarities between acoustic and GR BHs, the dynamics of these two systems are different, since the former are expected to satisfy hydrodynamical equations rather than Einstein's equations.

In this paper, we have investigated the viability of analogue gravity models for recovering the tidal properties of GR BHs, by studying their static tidal response. TLNs are usually introduced to describe the conservative response of compact objects to an external tidal field, and are found to vanish exactly for asymptotically flat GR BHs in four dimensional space-times, even though this property is fragile when going to higher dimensions.

We first studied the TLNs of (non)spinning (2+1) acoustic BHs, showing that in general they are non-vanishing and display a logarithmic running with radial distance, analogously to GR BHs in higher dimensions. We then extended the analysis to acoustic BHs in (3+1) dimensions, considering the canonical and de Laval nozzle metrics. We found that, contrary to the universal vanishing TLNs of Schwarzschild BHs in four dimensions, (3+1)-dimensional ABHs generally have non-vanishing response. The response runs logarithmically in the de Laval nozzle case, while taking vanishing values for canonical acoustic solutions only for certain angular multipole moments. 

Lastly, we have shown that the vanishing property of ABH TLNs for certain angular multipole moments can be understood through ladder symmetries of the perturbation equations, thus showing a strong similarity with GR BHs. Nonetheless, the general differences between these families of BHs confirm the expectation that the dynamics of acoustic BHs does not exactly reproduce the one of standard GR BHs. 

There are many directions of future research. An immediate goal is to generalize the computation of TLNs to spinning ABHs in four dimensions, in order to perform a detailed comparison with Kerr solutions in GR. Furthermore, it would be interesting to extend our analysis to dynamical perturbations on ABHs, by computing their dynamical TLNs and dissipative coefficients. Finally, on a more optimistic note, it would be fascinating to think about experimental setups to probe the tidal response of supersonic acoustic flows in laboratories. Exploring these extensions is left to future work. 

\vspace{0.3cm}
\section*{Acknowledgments}
\noindent
We thank A. Garoffolo for useful discussions, and especially J. Beltrán Jimenez, D. Bettoni and P. Brax for pointing out the existence of the small ladder structure.
V.DL. is supported by funds provided by the Center for Particle Cosmology at the University of Pennsylvania. B.K. acknowledges support from the National Science Foundation Graduate Research Fellowship under Grant No. DGE-2236662. The work of J.K. and M.T. is supported in part by the DOE (HEP) Award DE-SC0013528.

\bibliography{draft}

\end{document}